\documentclass[12pt]{iopart}

\usepackage{amssymb}
\usepackage{bm}
\usepackage[final,dvips]{epsfig}

\newcommand{\ff}[1]{{\bm #1}}
\newcommand{\ca}[1]{{\cal #1}}
\newcommand{\be}{\begin{equation}}
\newcommand{\ee}{\end{equation}}
\newcommand{\ba}{\begin{eqnarray}}
\newcommand{\ea}{\end{eqnarray}}
\newcommand{\ket}[1]{| #1 \rangle}
\newcommand{\bra}[1]{\langle #1 |}
\newcommand{\braket}[2]{\langle #1 | #2 \rangle}

\begin{document}

\title[]{Krylov-space approach to the equilibrium and the nonequilibrium single-particle Green's function}

\author{
Matthias Balzer 
\footnote{permanent address: Fraunhofer-Institut f\"ur Techno- und Wirtschaftsmathematik, Fraunhofer-Platz 1,
67663 Kaiserslautern, Germany}, 
Nadine Gdaniec 
\footnote{permanent address:
Universit\"at zu L\"ubeck, Institut f\"ur Signalverarbeitung, Ratzeburger Allee 160, 23562 L\"ubeck, 
and Philips Technologie GmbH Forschungslaboratorien, R\"ontgenstra\ss{}e 24-26, 22335 Hamburg, Germany} 
and Michael Potthoff}

\address{
I. Institut f\"ur Theoretische Physik,
Universit\"at Hamburg, 
Jungiusstra\ss{}e 9,
20355 Hamburg,
Germany
}

\ead{michael.potthoff@physik.uni-hamburg.de}

\begin{abstract}
The zero-temperature single-particle Green's function of correlated fermion models with moderately large Hilbert-space dimensions can be calculated by means of Krylov-space techniques. 
The conventional Lanczos approach consists of finding the ground state in a first step, followed by an approximation for the resolvent of the Hamiltonian in a second step.
We analyze the character of this approximation and discuss a numerically exact variant of the Lanczos method which is formulated in the time domain. 
This method is extended to get the nonequilibrium single-particle Green's function defined on the Keldysh-Matsubara contour in the complex time plane which describes the system's non-perturbative response to a sudden parameter switch in the Hamiltonian.
The proposed method will be important as an exact-diagonalization solver in the context of self-consistent or variational cluster-embedding schemes.
For the recently developed nonequilibrium cluster-perturbation theory, we discuss the efficient implementation and demonstrate the feasibility of the Krylov-based solver.
The dissipation of a strong local magnetic excitation into a non-interacting bath is considered as an example for applications. 
\end{abstract}


%\submitto{\JPCM}

% \maketitle
% 

\section{Introduction}
\label{intro}

Due to Wick's theorem, the single-particle Green's function $\ff G$ is the central quantity of interest in theoretical approaches to strongly correlated electron systems that are based on the concepts of weak-coupling perturbation theory \cite{AGD64,FW71}.
This holds for systems in thermal equilibrium as well as for systems subjected to strong time-dependent perturbations that give rise to a highly excited quantum state far from equilibrium \cite{Kel65,Dan84,Wag91}. 
Standard examples comprise the plain or different renormalized perturbation theories for lattice-fermion models like the Hubbard model \cite{SC91,PN97c}. 

Nonperturbative approximations can be constructed within dynamical variational approaches \cite{Pot05}. 
Here the single-particle Green's function or the self-energy is determined from a general variational principle.
This includes dynamical mean-field theory (DMFT) \cite{GKKR96,KV04}, different cluster extensions of the DMFT \cite{MJPH05}, the cluster-perturbation theory (CPT) \cite{SPPL00,GV93}, the variational cluster approach (VCA) \cite{Pot03a,PAD03} or the dual-fermion (DF) method \cite{RKL08}, for example.
These methods all involve a self-consistent or variational mapping onto an effective impurity or an effective cluster model for which $\ff G$ must be computed repeatedly. 
Among the standard ``solvers'' to treat those problems, such as the quantum Monte-Carlo method \cite{GML+11}, for example, exact diagonalization or the Lanczos technique represents an important alternative. 
Due to the exponential growth of the Hilbert space with increasing number of degrees of freedom, the Lanczos solver is basically restricted to single-band models and comparatively small clusters or single-site approximations with a few orbitals per site only. 

As suggested by Caffarel and Krauth \cite{CK94}, a two-step Lanczos procedure \cite{Lan50,LG93,Dag94,NM05} can be used as an efficient method to get the zero-temperature Green's function $\ff G$ of the single-impurity Anderson model:
After finding the approximate ground state of the model in a first Lanczos step, the frequency-dependent single-particle Green's function is obtained in the second step by approximating the resolvent $(\omega - H)^{-1} \mapsto \sum_m (\omega - E_m)^{-1} \ket{m}\bra{m}$ with the help of the eigenenergies $E_m$ and eigenstates $\ket{m}$ of the Hamiltonian $H$ in a small Krylov subspace of the full Hilbert space.
This Lanczos solver has turned out to be very efficient and reliable and is frequently used within DMFT, cluster DMFT and VCA, see Refs.\ \cite{GKKR96,CK94,KSG08,DAH+04} for examples.

Its extension to the general nonequilibrium situation is, however, an open issue and represents the main motivation of the present paper.
In detail our motivations are the following:

First, we note that there is a growing need for theoretical nonequilibrium approaches to describe and understand recent experimental studies.
This includes spin-relaxation and switching processes in nanostructured systems with itinerant and correlated electrons which are experimentally accessible by means of scanning-tunneling microscope techniques, for example \cite{LEL+10,KWCW11}. 
Another field of interest is given by fast-demagnetization processes probed by femtosecond optical excitations \cite{MWD+09} or the nonequilibrium electronic structure of strongly correlated transition-metal oxides which may be monitored by femtosecond pump-probe spectroscopies \cite{PLL+06,WPBC09}. 
Furthermore, there is an urgent need for theory to understand the nonequilibrium dynamics of highly excited fermionic states realized in correlated systems of ultracold atoms in optical lattices \cite{SGJ+10}.

Second, there are straightforward extensions of DMFT \cite{FTZ06,SM02} and its cluster variants, of the CPT \cite{BP11} and the DF method \cite{JLB+11} to correlated lattice models far from thermal equilibrium. 
Roughly speaking, these extensions are obtained when the theory is re-formulated on the Keldysh-Matsubara contour in the complex time plane.
Different solvers for the resulting effective nonequilibrium impurity or cluster problem have been employed:
An analytical approach based on the solution of a closed set of equations is available in the case of the Falicov-Kimball model only \cite{FTZ06}.
For Hubbard-type systems, diagrammatic weak-coupling \cite{SM02} and strong-coupling \cite{EW10} perturbative techniques can be used, or variants of the continous-time quantum Monte-Carlo technique \cite{GML+11}.
As exact-diagonalization solvers within nonequilibrium single-site or cluster methods, only {\em full} diagonalization procedures have been employed so far, namely for the nonequilibrium dual-fermion approach \cite{JLB+11} and the nonequilibrium CPT \cite{BP11}.
Opposed to a Krylov-space construction, the Green's function $\ff G$ is calculated here from its spectral or Lehmann representation using a basis of the full Hilbert space of the effective impurity or cluster model.
This limits the conveniently accessible system size to $L_c=6$ sites/orbitals only which must be regarded as crucial as the convergence with $L_c$ is known to be exponentially fast \cite{CK94}.

Third, in case of a sudden strong quench of a model parameter, i.e.\ far from thermal equilibrium, it is advisable to focus on real-time single-particle correlation functions of the form $\langle c_{\alpha}(t) c^\dagger_{\alpha'}(t') \rangle$.
Here $\alpha,\alpha'$ refer to one-electron orbitals and $\langle \cdots \rangle$ is the expectation value with an initial state different from the ground state or an eigenstate of the time-independent Hamiltonian $H$. 
In this case, the question for an approximation of the operator exponential $\exp(-iHt)$ rather than the resolvent must be addressed. 
A standard and reliable Krylov-space method is available to propagate a given state $|\Psi\rangle$ via $\exp(-iHt) \ket{\Psi}$ \cite{PL86,HL97,HL99,ML03,MMN05}.
The method is also used in the context of the density-matrix renormalization \cite{MMN05} and continuous-time quantum Monte-Carlo \cite{LW09}. 
In the present paper, we focus on an efficient application of this Krylov approach to get the Green's function on the Keldysh-Matsubara contour with a maximum real time $t_{\rm max}$ and a time discretization step $\Delta t$ typical for solver applications.
To demonstrate its feasibility, we employ the approach within the context of nonequilibrium CPT.

Finally, for the {\em equilibrium} single-particle Green's function $\ff G$, the Krylov construction to approximate the exponential $\exp(-iHt)$ represents an alternative approach to the conventional Lanczos technique which approximates the resolvent $(\omega - H)^{-1}$.
Their mutual relation shall be worked out here.

The paper is organized as follows:
The next section gives a brief overview of the Krylov construction and the standard Lanczos approach for the equilibrium Green's function. 
The approximate nature of the approach is made clear and the type of the approximation is characterized.
Section \ref{exact} discusses a variant of the technique formulated in the time domain by which the numerically exact Green's function is accessible.
In section \ref{nonegf} we seek an algorithm to get the nonequilibrium Green's function on the Keldysh-Matsubara contour in the complex time plane and propose a four-step Krylov-space based technique. 
To demonstrate its feasibility in the context of a cluster-embedding scheme, we consider the dissipation of a local magnetic excitation in section \ref{necpt} by means of nonequilibrium CPT.
Section \ref{con} summarizes the main results.

\section{Krylov space basis and equilibrium Green's function}
\label{lanc}

We start by giving a brief overview of the Lanczos approach to the single-particle Green's function. 
Details can be found in Ref.\ \cite{NM05}, for example.

For a given initial state $\ket{i_0}$, the $n$-th Krylov space is defined as
\begin{equation}
  \ca K_n(\ket{i_0}) = \mbox{span} \{ \ket{i_0}, H \ket{i_0}, ... , H^{n-1} \ket{i_0} \} \: ,
\label{kspace}
\end{equation}
where $H$ is the Hamiltonian of the system.
$\ca K_n(\ket{i_0})$ is an $n$-dimensional subspace of the full Hilbert space with dimension $d$. 
Usually, we consider $n \ll d$.
A basis of $\ca K_n(\ket{i_0})$ can be constructed by means of the numerically efficient recursion scheme
\begin{equation}
  \ket{i_{k+1}} = H \ket{i_k} - a_k \ket{i_k} - b_k^2 \ket{i_{k-1}} \qquad (k=0,...,n-1)
\label{recursion}
\end{equation}
with initial values $b_0 \equiv 0$ and $\ket{i_{-1}} \equiv 0$ and with the coefficients $a_k = \bra{i_k} H \ket{i_k} / \braket{i_k}{i_k}$ and $b_k^2 = \braket{i_k}{i_k} / \braket{i_{k-1}}{i_{k-1}}$.
Subsequent normalization yields the orthonormal Lanczos basis $\{ \ket{i_0}, ..., \ket{i_{n-1}}\}$ of $\ca K_n(\ket{i_0})$. 
In this basis, the Hamiltonian is represented by a tridiagonal matrix $\ff T$ with diagonal elements given by $a_0, ..., a_{n-1}$ and off-diagonal elements by $b_1, ... , b_{n-1}$:
Let $\ff H$ be the $d\times d$ matrix representation of the Hamiltonian in an arbitrary basis $\{ \ket{j} \}$, e.g.\ in the occupation-number basis where $\ket{j} = \ket{n_1,n_2,...,n_\alpha, ...}$ and $n_\alpha$ are the occupations of single-particle orbitals $\ket{\alpha}$. 
We have $H_{jj'} = \bra{j}H\ket{j'}$.
Let $\ff V = (\ff i_0 , ..., \ff i_{n-1})$ be the $d \times n$ matrix constructed from columns $\ff i_k$ representing $\ket{i_k}$ in the given basis, i.e.\ $V_{jk} =\braket{j}{i_k}$.
Then
\begin{equation}
  \ff T = \ff V^\dagger \ff H \ff V \: .
\label{tmatrix}
\end{equation}
Diagonalization of $\ff T$, 
\begin{equation}
  \ff D = \ff Q^\dagger \ff T \ff Q 
\label{tdiag}
\end{equation}
with a unitary $n\times n$ matrix $\ff Q$, yields a diagonal matrix $\ff D$ containing approximate eigenenergies $E_m$ of $H$.
The corresponding approximate eigenvectors, i.e.\ $H \ket{m} \approx E_m \ket{m}$ for $m=1,...,n$, are 
\begin{equation}
  \ket{m} = \sum_j U_{ji_m} \ket{j} 
\label{eigen}
\end{equation}
where we have defined the $d\times n$ matrix $\ff U = \ff V \ff Q$.

The convergence of the extremal eigenenergies with increasing $n$ is very fast.
To get the ground-state energy and the ground state itself, numerically almost exact results can be obtained with of the order of $n=100$ Lanczos iterations \cite{Dag94}.
The initial state $\ket{i_0}$ is arbitrary but must have a finite overlap with the ground state.

Consider now the single-particle Green's function. 
For frequency $\omega >0$, the zero-temperature retarded Green's function is given by 
\begin{equation}
  G_{\alpha\alpha'}(\omega) = G_{\alpha\alpha'}^{(>)}(\omega) 
  = 
  \langle 0 | c_\alpha \frac{1}{\omega + i \eta - H + E_0} c_{\alpha'}^\dagger | 0 \rangle
\label{gomega}
\end{equation}
where $\ket{0}$ is the ground state which is assumed to be nondegenerate, $E_0$ is the ground-state energy, $c_\alpha$ annihilates a fermion in the one-particle orbital $\ket{\alpha}$, and $\eta$ is a small positive number to shift the poles of the Green's function below the real axis in the complex frequency plane.
For $\omega < 0$, $G_{\alpha\alpha'}(\omega) = G_{\alpha\alpha'}^{(<)}(\omega) = \langle 0 | c_{\alpha'}^\dagger (\omega + i \eta + H - E_0)^{-1} c_{\alpha} | 0 \rangle$.

The Lanczos procedure \cite{Dag94,NM05} to get the Green's function consists in the following approximation for the resolvent (for $\omega>0$):
\begin{equation}
  \frac{1}{\omega + i \eta - H + E_0} \mapsto \sum_m
  \frac{1}{\omega + i \eta - E_m + E_0} \ket{m} \bra{m}
\label{resolv}
\end{equation}
where the $n \ll d$ approximate energy eigenstates $\ket{m}$ are obtained with Eq.\ (\ref{eigen}) from a second Lanczos run using $\ket{i_0} = c_{\alpha'}^\dagger \ket{0}$ as the initial state.
Typically, $n \sim 100$ is used again.
This yields a Lanczos Green's function with exactly $n$ poles:
\begin{equation}
  G_{\alpha\alpha'}(\omega) 
  \approx 
  G^{(L)}_{\alpha\alpha'}(\omega) 
  \equiv
  \sum_m \langle 0 | c_\alpha \ket{m} \frac{1}{\omega + i \eta - E_m + E_0} \bra{m} c_{\alpha'}^\dagger | 0 \rangle \: .
\label{gapprox}
\end{equation}
A continuous spectral function $A_{\alpha\alpha'}(\omega) = - (1/\pi) \mbox{Im} G_{\alpha\alpha'}(\omega)$ is obtained by Lorentzian broadening with a finite $\eta > 0$.

To estimate the quality of the approximation (\ref{resolv}), we consider the high-frequency expansion of the exact Green's function (\ref{gomega}) for $\eta \to 0$,
\begin{equation}
  G_{\alpha\alpha'}(\omega) 
  = 
  \sum_{r=0}^\infty \frac{1}{\omega^{r+1}} \langle 0 | c_\alpha (H - E_0)^r c_{\alpha'}^\dagger | 0 \rangle \; ,
\label{ghigh}
\end{equation}
and compare with the high-frequency expansion of the Lanczos Green's function (\ref{gapprox}).
Introducing $P \equiv \sum_m \ket{m} \bra{m}$ as the projector onto $\ca K_n(\ket{i_0})$ with $\ket{i_0} = c_{\alpha'}^\dagger \ket{0}$, we immediately get for the latter:
\begin{equation}
  G^{(L)}_{\alpha\alpha'}(\omega) 
  = 
  \sum_{r=0}^\infty \frac{1}{\omega^{r+1}} \langle 0 | c_\alpha (P(H - E_0)P)^r c_{\alpha'}^\dagger | 0 \rangle \: .
\label{glhigh}
\end{equation}
Here, we have assumed that the error in the determination of the ground state $\ket{0}$ can be neglected. 
This is usually an excellent approximation which will also be adopted in the rest of the paper.
Comparing Eqs.\ (\ref{ghigh}) and (\ref{glhigh}) shows that the Lanczos approximation for the Green's function conserves the first $n$ coefficients in the high-frequency expansion since $(H - E_0)^r c_{\alpha'}^\dagger | 0 \rangle = (H - E_0)^r | i_0 \rangle = (P(H - E_0)P)^r | i_0 \rangle \in \ca K_n(\ket{i_0})$ if $r \le n-1$.

The expansion coefficients determine the first $n$ moments $\int_{-\infty}^\infty d\omega \, \omega^r A_{\alpha\alpha'}(\omega)$ of the spectral function. 
Therefore, we can conclude that the Lanczos technique at iteration depth $n$ provides a spectral function with the correct first $n$ moments --- irrespective of the fact that the excited states $\ket{m}$ are obtained with a much lower accuracy than the ground state.
However, significant deviations from the exact spectral function are expected at high excitation energies since the convergence with increasing $n$ is known to be faster for low-lying as compared to highly excited states.

\section{Numerically exact computation of the Green's function}
\label{exact}

Seeking for an improved approximation, let us consider the time-dependent Green's function
\begin{equation}
  G_{\alpha\alpha'}(t) = \frac{1}{2\pi} \int_{-\infty}^\infty d\omega \: 
                         e^{-i\omega t} G_{\alpha\alpha'}(\omega) 
\label{ft}
\end{equation}
which is obtained from the frequency-dependent Green's function in Eq.\ (\ref{gomega}) via Fourier transformation. 
A straightforward calculation yields $
  G_{\alpha\alpha'}(t) = - i \Theta(t) \, \bra{0} c_\alpha e^{-i (H-E_0) t} 
  c_{\alpha'}^\dagger \ket {0}e^{-\eta t}$.
At this point we can employ a Krylov-space technique \cite{PL86,HL97,HL99,ML03,MMN05} to compute the time evolution of the state
\begin{equation}
  \ket{\Psi_{\alpha'}(t)} \equiv e^{-i H t} c_{\alpha'}^\dagger \ket {0} \: ,
\label{psit}
\end{equation}
and therewith 
\begin{equation}
G_{\alpha\alpha'}(t) = - i \Theta(t) \, \bra{0} c_\alpha \ket{\Psi_{\alpha'}(t)} e^{i (E_0 +i \eta) t} \: .
\label{psigreen}
\end{equation}

The idea is the following:
If $\Delta t$ is sufficiently small, the Taylor expansion of the operator exponential $\exp(-iH\Delta t)$ can be truncated at some finite small order $n \ll d$ within numerical accuracy.
This implies that the state $\ket{\Psi_{\alpha'}(t+\Delta t)} = \exp(-i H \Delta t) \ket{\Psi_{\alpha'}(t)}$ lies in the Krylov space $\ca K_n(t)$ constructed from the initial state $\ket{i_0} = \ket{ \Psi_{\alpha'}(t)}$ at time $t$ (for simplicity we suppress the $\alpha'$ dependence of $\ca K_n(t)$ in the notation).
Hence, 
\begin{equation}
  \ket{\Psi_{\alpha'}(t+\Delta t)} = \exp(-i P(t)HP(t) \Delta t) \ket{\Psi_{\alpha'}(t)} \: ,
\label{newstate}
\end{equation}
where $P(t)$ is the projector onto $\ca K_n(t)$.
Within this Krylov space, the time evolution operator can be represented as 
\begin{equation}
  \exp(-i P(t)HP(t) \Delta t) 
  = 
  \sum_{jj'} \ket{j} [\exp(-i \ff V(t) \ff T(t) \ff V^\dagger(t)\Delta t)]_{jj'} \bra{j'} \: ,
\end{equation}
where $\ff V(t)$ and $\ff T(t)$ are the representations of the Lanczos basis and the Hamiltonian obtained during the construction of $\ca K_n(t)$ via the Lanczos iteration.
Using the orthonormality of the Lanczos basis, $\ff V^\dagger \ff V=\ff 1 \ne \ff V \ff V^\dagger$, and Eq.\ (\ref{tdiag}), we have
\begin{equation}
  \exp(-i P(t)HP(t) \Delta t) 
  = 
  \sum_{jj'} \ket{j} [\ff U(t)
  \exp(-i \ff D(t) \Delta t) 
  \ff U^\dagger(t)]_{jj'} \bra{j'} \: ,
\label{udelta}
\end{equation}
where $\ff D(t) = \ff Q(t)^\dagger \ff T(t) \ff Q(t)$ and $\ff U(t) = \ff V(t) \ff Q(t)$.

For sufficiently short $\Delta t$, Eq.\ (\ref{udelta}) thus provides a {\em numerically exact} way to propagate the state (\ref{psit}) by $\Delta t$ using the Lanczos recursion algorithm.
The time propagation can be repeated by restarting the algorithm with the state at $t+\Delta t$ as the new initial state. 
Using several restarts, this allows us to compute the time-dependent Green's function from $t=0$ up to a time $t_{\rm max}$ at which, depending on the choice of $\eta$, the exponential damping $e^{-\eta t}$ ensures convergence of the time integral in the Fourier back transformation. 
This yields the frequency-dependent Green's function $G_{\alpha\alpha'}(\omega)$ for a given $\eta$.
It is worth mentioning that this approach (``time-dependent Lanczos'') is numerically exact, opposed to the ``conventional'' Lanczos procedure described in the preceding section. 
The essential difference is that the computation of the operator exponential $\exp(-i H  t)$ can be decomposed into several steps with short $\Delta t$, and that $\exp(-iH\Delta t)$ can be represented numerically exactly by means of a low-dimensional Krylov space.
On the other hand, the resolvent $1/(\omega - H)$ must be computed in a single step and thus be represented in a single (larger) Krylov space. 

%*******************************************************************************
\begin{figure}[t]
\centerline{\includegraphics[width=0.9\columnwidth]{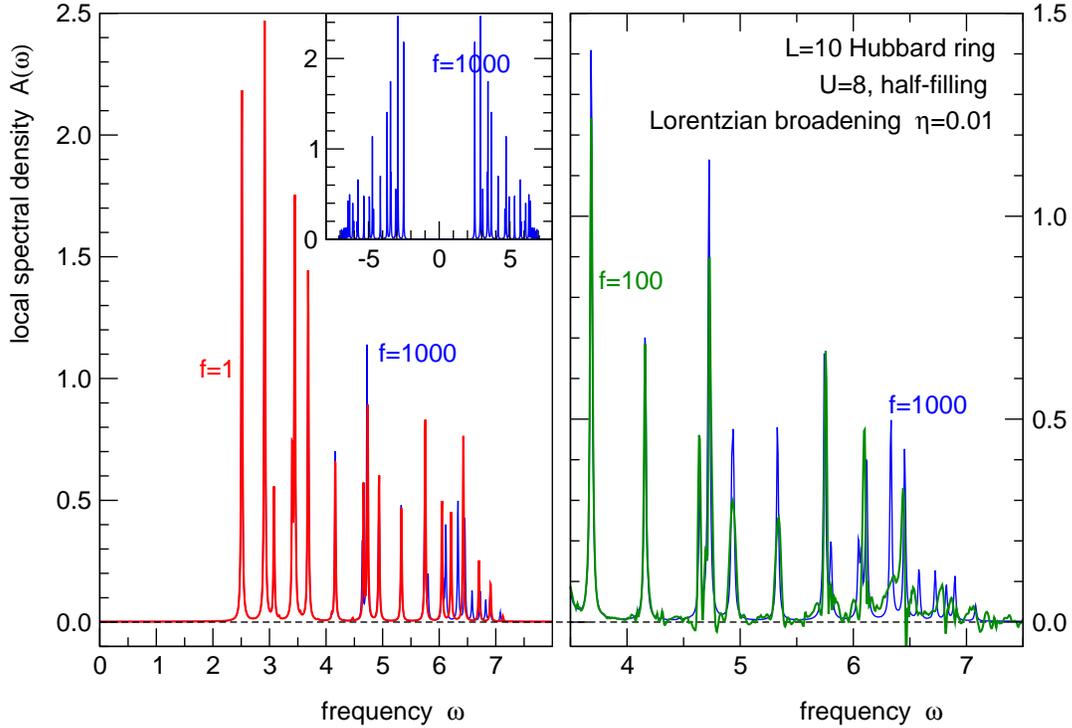}}
\caption{ 
Local spectral density $A(\omega) = (-1/\pi) \mbox{Im} \, G_{ii}(\omega)$ of the particle-hole symmetric Hubbard model on an $L=10$ site ring at $U=8$. The nearest-neighbor hopping $T=1$ sets the energy scale. Lorentzian broadening: $\eta=0.01$, maximum time for propagation of states: $t_{\rm max}=1000.0$. 
{\em Inset:} numerically exact spectral function obtained with $f=1000$ Lanczos restarts, i.e.\ $\Delta t=1.0$.
{\em Left panel:} $f=1$, $\Delta t = 1000.0$ (red fat line) compared with $f=1000$, $\Delta t=1.0$ (blue thin line).
{\em Right panel:} $f=100$, $\Delta t =10.0$ (green fat line) compared with $f=1000$, $\Delta t=1.0$ (blue thin line).
}
\label{equigf}
\end{figure}
%*******************************************************************************

Figure \ref{equigf} gives an example for the one-dimensional Hubbard model
\begin{equation}
  H = - T \sum_{\langle ij\rangle} \sum_{\sigma=\uparrow,\downarrow} c_{i\sigma}^\dagger c_{j\sigma}
  + U \sum_i n_{i\uparrow} n_{i\downarrow} \: 
\end{equation}
with nearest-neighbor hopping $T=1$ and Hubbard interaction $U=8$ at half-filling.
We consider a system with $L=10$ sites and periodic boundary conditions. 
The local retarded Green's function $G_{ii}(\omega)$ is calculated via numerical fast Fourier transformation from the time-dependent Green's function $G_{ii}(t)$. 
The latter is obtained via Eq.\ (\ref{psigreen}) for times up to $t_{\rm max} = 1000.0$ which is sufficient to ensure the convergence of the Fourier transform at a Lorentzian broadening of $\eta = 0.01$.

For the propagation of a state by means of the Krylov technique, Eq.\ (\ref{psit}), it is more advantageous to consider a large Krylov space dimension $n$ and a large time propagation step $\Delta t$ but a smaller number of restarts $f$ opposed to a small $n$ and short $\Delta t$ but more restarts $f$. 
On the other hand, the Krylov space dimension should not be much larger than $n = \ca O(100)$ since all states of the Lanczos basis (i.e.\ $\ff V$) have to be stored.
For the present calculation, we fix $n=100$. 
To avoid a loss of orthogonality of the Lanczos basis states during the iterative procedure, a Gram-Schmidt reorthogonalization scheme is employed.

The spectrum obtained with $f=1000$ restarts, corresponding to $\Delta t = 1.0$, is shown in the inset of figure \ref{equigf}. 
This represents the numerically exact solution, any increase of $f$ or $n$ does not change the results.
Note that Eq.\ (\ref{udelta}) can be used to compute the Green's function on a finer time grid with spacing $\Delta t' < \Delta t$ without additional restarts. 
Here we have used $\Delta t' = 0.01$ independent of the different $f$ considered.
In the left panel of the figure the numerically exact result is compared with the spectrum obtained for $f=1$ (due to particle-hole symmetry only frequencies $\omega>0$ are displayed).
Here, the Krylov space is constructed only once, i.e.\ $\Delta t = t_{\rm max}= 1000.0$.
As can be seen from the figure, there is a perfect agreement for lower frequencies while deviations are clearly visible for the number and the energy position of the peaks as well as for their spectral weights at higher frequencies.

It is important to realize that the $f=1$ spectrum just corresponds to the result of the conventional Lanczos method:
If the Krylov space is constructed only once from the initial state $\ket{i_0} = c_{\alpha'}^\dagger \ket{0}$, the matrices $\ff D$ and $\ff U$ in Eq.\ (\ref{udelta}) are independent of $t$. 
Inserting Eq.\ (\ref{udelta}) with $\Delta t$ replaced by $t$ into Eq.\ (\ref{newstate}) and Eq.\ (\ref{psigreen}), yields
\begin{equation}
G_{\alpha\alpha'}(t) = - i \Theta(t) \, \sum_m \bra{0} c_\alpha \ket{m} e^{-iE_m t} \bra{m} c_{\alpha'}^\dagger \ket{0} e^{i (E_0 +i \eta) t} \: ,
\label{res}
\end{equation}
where $m$ just runs over the approximate eigenstates, see Eq.\ (\ref{eigen}), that are obtained from the conventional Lanczos technique. 
Fourier transformation of Eq.\ (\ref{res}) gives $G^{(L)}_{\alpha\alpha'}(\omega)$ as defined in Eq.\ (\ref{gapprox}).
We arrive at the conclusion that time-dependent Lanczos carried out with a single Krylov space ($f=1$, see figure \ref{equigf}) is equivalent with the conventional Lanczos method. 
This has also been checked numerically.

The left panel of figure \ref{equigf} therefore shows the deviations of the conventional Lanczos method from the exact result. 
The perfect agreement at low frequencies is now easily explained by the fact that the ground state and the low-lying excited eigenstates of $H$ are accurately predicted by the conventional Lanczos method. 
Discrepancies at higher frequencies of the order of $U$ are attributed to the poor convergence of higher excited states. 
At even higher excitation energies, outside the support of the spectrum, the conventional Lanczos Green's function becomes reliable again since the first $n$ moments and thus the corresponding coefficients in the high-frequency expansion are predicted correctly as discussed in section \ref{lanc}.
Note that the time-dependent Lanczos approach does not make any reference to the excited eigenstates of $H$ although for a large Krylov-space dimension, such as $n=100$, the elements of $\ff D(t)$ and of $\ff U(t)$ may be close to the eigenenergies and to the coefficients of the eigenstates [Eq.\ (\ref{eigen})] and only weakly dependent on the initial state for the Lanczos restart at time $t$. 
On the other hand, this weak dependence on the initial state is important to get the numerically exact result.

The time-dependent Lanczos is as memory efficient as the conventional one. 
Since all states of the Lanczos basis have to be stored (the matrix $\ff V$), memory requirements are minimized for a small $n$. 
Very small Krylov-space dimensions (e.g.\ $n<10$) may be used at the cost of an increased number of restarts $f$ (i.e.\ short $\Delta t$).
On the other hand, CPU time is minimized with a small $f$ and large $n$. 
As compared to the conventional method, the computational cost is to a very good approximation higher by a factor $f$ (for the same $n$) since the Krylov space must be constructed $f$ times.
This must be kept in mind for applications like DMFT. 

This raises the question for a possible compromise: 
Can a {\em small} number of restarts $f$ cure the errors of the conventional Lanczos approach?
The right panel of figure \ref{equigf} shows the result of the spectral function from a calculation with only $f=100$ restarts keeping the Krylov-space dimension unchanged ($n=100$). 
It turns out that the time evolution of the state, Eq.\ (\ref{newstate}), is no longer exact over the entire time interval $\Delta t = 10.0$ for each restart.
Compared to the conventional Lanczos method, the deviations from the exact spectral function are of the same order of magnitude. 
More important, however, the approximation is no longer causal and produces negative spectral weight as can be seen from the figure. 
The case $f=1$ represents an exception. 
Here, Eq.\ (\ref{gapprox}) applies and the non-negativity of the local spectral function is obvious.

\section{Nonequilibrium Green's function}
\label{nonegf}

The nonequilibrium single-particle Green's function depends on {\em two} time arguments and is given by 
\begin{equation}
  G_{\alpha\alpha'}(z,z') = - i \langle 0 | \ca T c_\alpha(z) c_{\alpha'}^\dagger(z') | 0 \rangle
\label{gnon}
\end{equation}
where $\ket{0}$ is an arbitrary state, usually not an eigenstate of $H$, which describes the system at a time $t=0$, and where the annihilator and the creator are given in the Heisenberg picture with times $z,z'$ on the Keldysh-Matsubara contour in the complex time plane \cite{Wag91}. 
$\ca T$ denotes the time ordering on the contour.
We are seeking an algorithm to compute the Green's function by means of a Krylov-space technique that meets the requirements for a ``solver'' in the context of nonequilibrium DMFT \cite{FTZ06,EKW09} or cluster-embedding approximations, such as the nonequilibrium CPT \cite{BP11}.
This means that impurity or cluster models at half-filling with more than $L=6$ sites should be accessible.
In any application as a solver, Dyson's equation, which is an integral equation on the contour, must be solved by time discretization and numerical matrix inversion.
Therefore, the Green's function must be computed on a discrete time mesh on the contour that is sufficiently fine for applications of standard quadrature formulas.
Finally, since inversions of matrices in the time variables are involved, the typical maximum time $t_{\rm max}$ up to which the time propagation of observables is traced is comparatively small, e.g.\ $t_{\rm max} = 10$ in units of the inverse hopping.

Let us briefly discuss the possibility to compute the Green's function in frequency space where, like in the conventional Lanczos approach, the resolvent is approximated. 
Consider, for example, the lesser Green's function
\begin{equation}
  G^{<}_{\alpha\alpha'}(t,t') = i \langle 0 | c_{\alpha'}^\dagger(t') c_\alpha(t) | 0 \rangle
  = i 
  \langle 0 | e^{iH t'} c_{\alpha'}^\dagger e^{iH(t-t')} c_\alpha e^{-iHt} | 0 \rangle
\label{glesser}
\end{equation}
with real time arguments $t,t'>0$ and $t-t'>0$. 
Using the identity
$\int_{-\infty}^\infty d\omega e^{-i\omega t} (\omega+i\eta-H)^{-1} = - 2\pi i \Theta(t) e^{-iHt} e^{-\eta t}$,
this can be written as
\begin{eqnarray}
  G^{<}_{\alpha\alpha'}(t,t') 
  &=&  
  \frac{1}{(2\pi)^3} \int\!\!\int\!\!\int d\omega_1d\omega_2d\omega_3\:
  e^{-i\omega_1t'}
  e^{-i\omega_2(t-t')}
  e^{-i\omega_3t} \times
\nonumber \\
  &\times&
  \langle 0 | 
  \frac{1}{\omega_1 +i\eta+H} 
  c_{\alpha'}^\dagger
  \frac{1}{\omega_2 +i\eta+H} 
  c_\alpha
  \frac{1}{\omega_3 +i\eta-H} 
  | 0 \rangle \: .
\label{glesseromega}
\end{eqnarray}
Now, using $\ket{0}$ as the initial state for the Lanczos iterations, one may construct the Krylov space $\ca K_n(\ket{0})$ and, using Eq.\ (\ref{eigen}), the orthonormal basis $\{\ket{m}\}$ of $\ca K_n(\ket{0})$ consisting of approximate eigenstates of $H$.
Therewith the resolvents $(\omega_3 +i\eta-H)^{-1}$ and $(\omega_1 +i\eta+H)^{-1}$ can be approximated like in Eq.\ (\ref{resolv}).
For the remaining resolvent $(\omega_2 +i\eta+H)^{-1}$, another basis must be constructed for any $m$ if, in the spirit of the conventional Lanczos approach, $c_\alpha \ket{m}$ shall be used as the respective initial state.
Even then, however, the high-frequency asymptotics cannot be recovered correctly, opposed to the equilibrium case.
This final step, therefore, represents a crude approximation.

A numerically exact access to $\ket{\Psi(\omega_3)} \equiv (\omega_3 +i\eta-H)^{-1} | 0 \rangle$ (and likewise to 
$\langle 0 | (\omega_1 +i\eta +H)^{-1}$) would be provided by the correction-vector method \cite{SR84,SR89} which is frequently employed in the context of dynamical density-matrix renormalization \cite{RPK+97,PRSB99,KW99,Jec02}. 
For each frequency $\omega_3$, the correction vector $\ket{\Psi(\omega_3)}$ can be obtained as the solution of a sparse inhomogeneous system of linear equations $(\omega_3 +i\eta-H) \ket{\Psi(\omega_3)} = | 0 \rangle$ with a dimension given by the Hilbert-space dimension. 
To evaluate Eq.\ (\ref{glesseromega}), however, another correction vector, depending on two frequency arguments, must be computed as the solution of $(\omega_2 +i\eta+H) \ket{\chi(\omega_2,\omega_3)} = c_\alpha | \Psi(\omega_3) \rangle$. 
This appears as less efficient than approaches working in the time domain directly.

We therefore propose the following four-step procedure to compute the $t,t'$-dependent Green's function:

(i) The system's initial state $\ket{0}$ must be given or is calculated by means of a standard Lanczos procedure as the ground state of an initial-state Hamiltonian $H_{\rm ini} \ne H$.

(ii) Constructing the Krylov space with $\ket{0}$ as the initial state for the Lanczos iteration and, depending on $n$ and $t_{\rm max}$, using $f$ additional restarts, the state
\be
\ket{\Phi(t)} \equiv e^{-iHt} \ket{0}
\ee
is computed with numerical accuracy and stored on a discrete time mesh for all times up to $t_{\rm max}$.

(iii) For each orbital $\alpha$ of interest, the state
\be
\ket{\Psi_\alpha(t)} \equiv
e^{iHt} c_\alpha \ket{\Phi(t)}
\ee
is computed with numerical accuracy and stored on the time mesh for all $t$ up to $t_{\rm max}$. 
This step is most time consuming since the Krylov time evolution must be performed for any $t$ on the time mesh, i.e.\ for any initial state $c_\alpha \ket{\Phi(t)}$.
The CPU time for the construction of a Krylov space scales linearly with the dimension $n$. 
While usually it is efficient to employ large Krylov spaces and longer propagation times $\Delta t$, one has to bear in mind that the Green's function (\ref{gnon}) must be obtained on a fine time mesh which requires, see Eq.\ (\ref{udelta}), the computation of $\ca O (n)$ dot products of Hilbert-space vectors for each time on the mesh. 
Therefore, it is not advisable to construct too large Krylov spaces.
With increasing $n$, the CPU time is eventually dominated by the evaluation of Eq.\ (\ref{udelta}) rather than by the Krylov-space construction for the case that there are many time points within a single interval $\Delta t$.

% note: the time evolution is done via:
% 
% - $\ff V \ff U \exp(-i\ff D \Delta t) \ff U^\dagger \ff V^\dagger \ket{ini}$ 
% 
% - $\ff V \ff U \exp(-i\ff D \Delta t) \ff U^\dagger \ket{1,0,0,0,0}$ 
% 
% - $\ff V \ff U \exp(-i\ff D \Delta t) \ket{\mbox{first row}}$ 
% 
% - $\ff V \times \ff v(t)$ wit $\ff v$ small vector (100 dimensions), time dependent

(iv) Finally, the lesser Green's function at arbitrary times $t,t'$ is obtained as the scalar product:
\be
G^{<}_{\alpha\alpha'}(t,t') = i \braket{\Psi_{\alpha'}(t')}{\Psi_\alpha(t)} \: .
\ee
Likewise $G^{>}_{\alpha\alpha'}(t,t')$ and Green's functions $G^\rceil_{\alpha\alpha'}(t,\tau')$ and $G^\lceil_{\alpha\alpha'}(\tau,t')$ with mixed real/imaginary time arguments can be calculated while the Matsubara Green's function $G^{\rm M}_{\alpha\alpha'}(\tau,\tau') = G^{\rm M}_{\alpha\alpha'}(\tau-\tau')$ is the equilibrium Green's function for the initial-state Hamiltonian $H_{\rm ini}$ and accessible by the time-dependent or by the conventional Lanczos method described in section \ref{nonegf}.

%*******************************************************************************
\begin{figure}[t]
\centerline{\includegraphics[width=0.8\columnwidth]{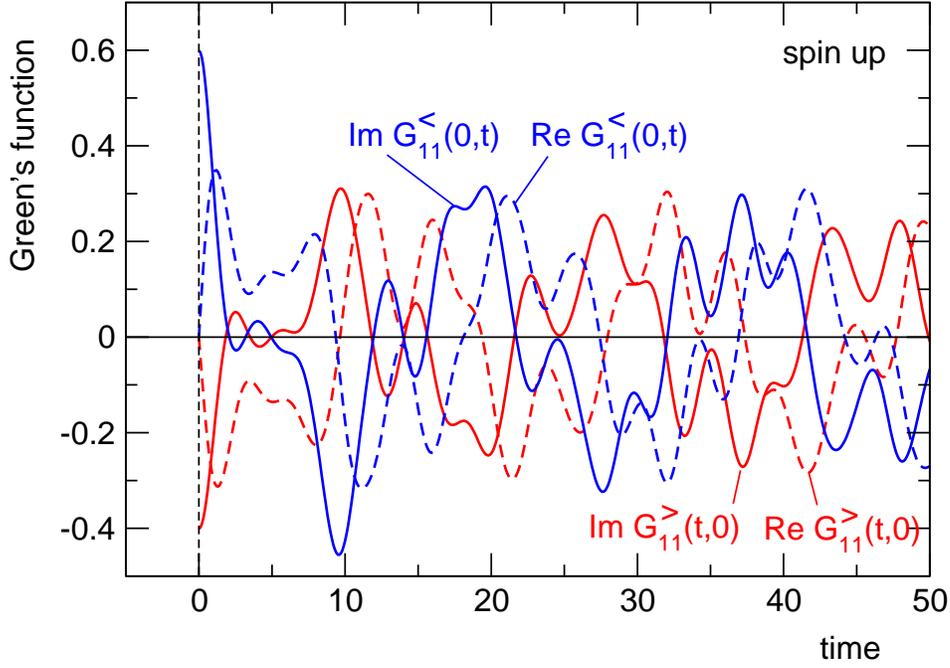}}
\caption{
Lesser and greater local spin-$\uparrow$ Green's function $G^<_{11}(0,t)$ and $G^>_{11}(t,0)$ as functions of $t$ for a single-impurity particle-hole symmetric Anderson model with $L=8$ sites in a chain geometry with the first site ($i=1$) as the correlated impurity and with $U=1$.
Energy and time units are set by the nearest-neighbor hopping $T=1$.
At $t=0$ the system is prepared in a state that is given by the ground state of the same model but in the presence of a local magnetic field of strength $h=0.2$ applied to the impurity site. 
The field is switched off for $t>0$.
}
\label{gsiam}
\end{figure}
%*******************************************************************************
 
Figure \ref{gsiam} gives an example for the single-impurity particle-hole symmetric Anderson model on a one-dimensional chain with $L=8$ sites and with the correlated impurity at site 1: 
\be
  H = \varepsilon_{\rm imp} \sum_{\sigma} c_{1\sigma}^\dagger c_{1\sigma} 
    + U \sum_\sigma n_{1\uparrow} n_{1\downarrow}
    - T \sum_{i=1}^{L-1} \sum_\sigma \left( c_{i\sigma}^\dagger c_{i+1\sigma} + \mbox{H.c.} \right) \: .
\ee
Here, $T$ is the nearest-neighbor hopping, $U$ the Hubbard interaction, $\varepsilon_{\rm imp}=-U/2$, and $n_{1\sigma} = c_{1\sigma}^\dagger c_{1\sigma}$, $\sigma=\uparrow, \downarrow$.
The system's state $\ket{0}$ at time $t=0$ is defined as the ground state of an initial-state Hamiltonian $H_{\rm ini} = H + H_{\rm field}$ which includes a finite local magnetic field at the impurity site in addition. 
The field disturbs the system at $t=0$ and is switched off for $t>0$:
\be
  H_{\rm field}(t) = - h \Theta(-t) (n_{1\uparrow} - n_{1\downarrow}) \: .
\label{hfi}
\ee
The initial state $\ket{0}$ is calculated by means of the standard Lanczos technique with $n=200$. 
For $t=0$, we have $\mbox{Im} G^<_{11}(0,0) = \langle 0 | c^\dagger_{1\uparrow} c_{1\uparrow} | 0 \rangle$ and $\mbox{Im} G^>_{11}(0,0) = - \langle 0 | c_{1\uparrow} c^\dagger_{1\uparrow} | 0 \rangle$ for the spin-$\uparrow$ lesser and greater Green's functions, respectively.
Their difference is unity as can be seen in the figure and as required by the canonical anticommutator relations.
The real parts must vanish. 
For $t>0$, $G^<_{11}(0,t)$ and $G^>_{11}(t,0)$ become complex and show strong oscillations as it is typical for a finite-size system. 

We have compared the suggested Krylov-space method to compute the Green's function (\ref{gnon}) on the complete Keldysh-Matsubara contour with a full-diagonalization approach. 
With the latter all eigenstates of the Hamiltonian are obtained by numerical diagonalization. 
Time dependencies and $G_{\alpha\alpha'}(z,z')$ are easily obtained then.
Typically, the full-diagonalization approach is faster up to $L=6$ sites at half-filling and exploiting the symmetries due to conservation of the total particle number and the $z$-component of the total spin.
For $L = 8$ and larger systems, the Krylov approach is superior. 
Eventually, both the full diagonalization and the Krylov method are limited by the need to store matrices of size $d \times d$ or $d \times n$, respectively.
At half-filling $L=12$ sites are easily accessible with the Krylov method on a standard PC.
CPU times are an order of magnitude longer as compared to the standard Lanczos technique for the equilibrium Green's function.

\section{Nonequilibrium cluster-perturbation theory}
\label{necpt}

The nonequilibrium cluster-perturbation theory (NE-CPT) \cite{BP11} is a simple cluster-embedding approach and constructed as a straightforward generalization of the standard (equilibrium) CPT \cite{SPPL00,GV93}. 
The NE-CPT can in principle be applied to an arbitrary lattice model of correlated electrons with local interactions. 
The main idea is to partition the original lattice into smaller pieces (clusters) for which the Green's function can be computed by means of an exact-diagonalization approach. 
The Green's function $\ff G$ of the original model is then obtained from the cluster Green's function $\ff G'$, or more precisely the Green's function $\ff G'$ for the system of disconnected clusters, via the CPT equation:
\be
  \ff G = \frac{1}{{\ff G'}^{-1} - \ff V} \: .
\label{cpteq}
\ee
Here, $\ff V$ is the {\em inter}-cluster hopping. 
The CPT equation can be interpreted as a resummation of diagrams in a perturbative expansion of $\ff G$ around the limit of disconnected and non-interacting clusters \cite{BP11}.
Thereby certain vertex corrections are neglected which describe the effects of inter-cluster potential scattering on the electron self-energy. 
In the diagrammatic formulation, the step from the CPT to the NE-CPT is particularly clear since standard perturbation theory for a nonequilibrium situation basically follows just along the lines of perturbation theory for systems in thermal equilibrium \cite{Wag91}. 
As concerns the CPT, the essential new point is that all quantities in CPT equation (\ref{cpteq}) have to be interpreted as given on the Keldysh-Matsubara contour in the complex time plane, see equation (\ref{gnon}), and that the matrix inverse in (\ref{cpteq}) not only refers to the orbital indices but also to the time variables, i.e.\ $\ff G$ is actually given by the solution of an integral equation. 
While the conventional Lanczos method is frequently used as a solver for equilibrium CPT \cite{SPPL00,ZEAH00,SPP02,HAvdL05,MJPH05}, only full diagonalization has been considered for NE-CPT so far. 
Half-filled Hubbard clusters of no more than $L_c=6$ sites can thereby be treated conveniently. 

%*******************************************************************************
\begin{figure}[t]
\centerline{\includegraphics[width=0.6\columnwidth]{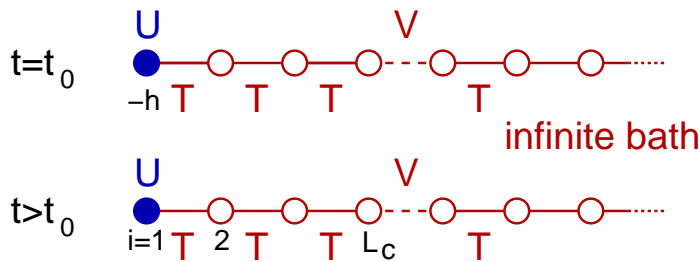}}
\caption{ 
Pictorial representation of the final-state Hamiltonian $H$ governing the time evolution (bottom) and the Hamiltonian $H_{\rm ini}$ generating the initial state as its ground state (top). 
Note that the Hubbard interaction $U$ is spatially separated from the hopping perturbation $V$ which links a small cluster of $L_c \le 12$ sites with an infinite uncorrelated bath in a semi-infinite chain geometry.
The initial state is defined to be the ground state in the presence of a finite local magnetic field $h=0.2$ applied to the impurity site.
The field is suddenly switched off at $t_0 = 0$. 
We consider the particle-hole symmetric model at half-filling.
}
\label{siam}
\end{figure}
%*******************************************************************************

Usually, the disregard of the mentioned vertex corrections represents a severe cluster mean-field-type approximation. 
The correction to the self-energy of lowest order in $V$, however, depends on the square and higher powers of the free {\em off-diagonal} Green's function that links sites with finite $U$ and $V$ interactions. 
In cases where these interactions are spatially separated, vertex corrections are expected to be small.
We therefore consider a single-impurity Anderson model (SIAM) on a semi-infinite chain with the impurity on the first site but with the ``inter-cluster'' hopping $V$ between sites $L_c$ and $L_c+1$ (see section \ref{nonegf} and figure \ref{siam}).
$V$ connects a small SIAM with $L_c \le 12$ sites and an infinite uncorrelated bath. 
The Green's function for the disconnected system with $V=0$ therefore consists of two independent parts: 
the Green's function of the isolated cluster and the bath Green's function. 

The bath Green's function $\ff G^{\rm (b)}$ is readily obtained as 
\begin{equation}
% \langle c_\alpha(t) c^\dagger_{\alpha'}(t') \rangle
G^{\rm (b)}_{ii'}(z,z') 
=
-i
\left( 
\frac{e^{-i \ff T_{\rm b} (z-z')} 
}{1+e^{-\beta \ff T_{\rm b} }} 
\right)_{ii'}
\label{gb1}
\end{equation}
if $z$ later than $z'$ on the contour and
\begin{equation}
% \langle c^\dagger_{\alpha'}(t') c_\alpha(t) \rangle
G^{\rm (b)}_{ii'}(z,z') 
= 
i
\left( 
\frac{e^{-i \ff T_{\rm b} (z-z')}}{e^{\beta \ff T_{\rm b}}+1}
\right)_{ii'}
\label{gb2}
\end{equation}
if $z'$ later than $z$.
Here, the parameter $\beta \to \infty$ projects out the ground state of the bath Hamiltonian with hopping matrix $\ff T_{\rm b}$.

%*******************************************************************************
\begin{figure}[t]
\centerline{\includegraphics[width=0.99\columnwidth]{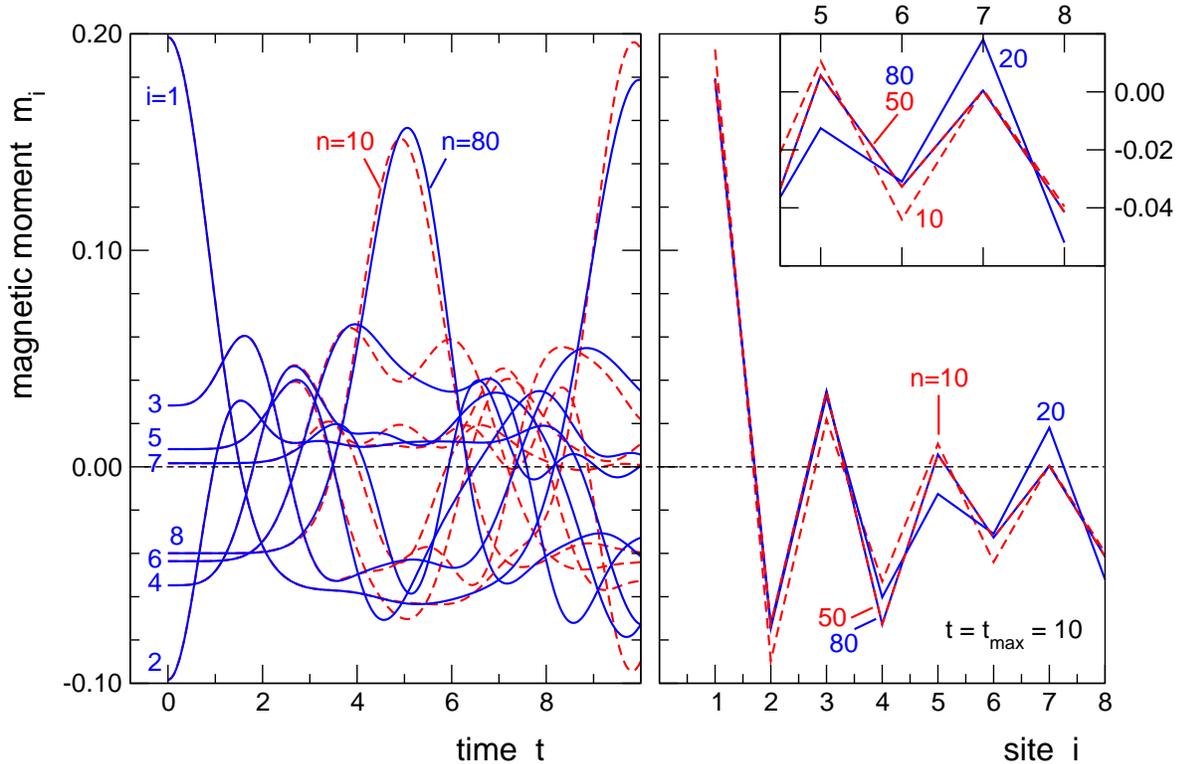}}
\caption{ 
{\em Left:}
Local magnetic moments $m_i = n_{i\uparrow} - n_{i\downarrow}$ in the initial state at $t=0$ and as functions of time $t>0$ at the cluster sites $i=1,...,L_c=8$ for the decoupled ($V=0$) model displayed in figure \ref{siam}.
{\em Right:}
Profile of the moment at $t=10$. 
{\em Inset:} $m_i$ for $i\ge 5$ on a larger scale.
Results are obtained from the nonequilibrium Green's function, $n_{i\sigma}(t) = -i G^<_{ii\sigma}(t,t)$, which has been calculated using the four-step Krylov-space method.
Parameters: $L_c=8$, $T=V=U=1$, $h=0.2$ for $t=0$, $h=0$ for $t>0$.
The initial state has been obtained as the ground state of $H_{\rm ini}$ using the conventional Lanczos technique with $n=200$ and using Gram-Schmidt reorthogonalization. 
For the final-state dynamics $\Delta t = t_{\rm max} =10$ (no restart) and different Krylov-space dimensions (as indicated) have been used. 

}
\label{mag}
\end{figure}
%*******************************************************************************

The Green's function of the isolated cluster is calculated using the Krylov-space method discussed in the preceding section.
Its dependence on the time variables is inhomogeneous and reflects the system's time evolution after an initial perturbation.
The initial state is taken to be the ground state of the system but with a finite local magnetic field applied to the impurity site with strength $h=0.2$, see equation (\ref{hfi}).
This field polarizes the vicinity of the impurity site in the finite cluster.
Figure \ref{mag} gives an example for a cluster size $L_c=8$.
At $t=0$ there is a strong local magnetic moment at the impurity site. 
With increasing distance from the impurity the moments alternate around zero and decrease in size.
Note that the total polarization $\sum_{i=1}^{L_c} m_i = 0$ as the field is too weak to break up the singlet ground state of the cluster.

The field is suddenly switched off at $t=0$. 
The site-dependent moments for times $t>0$ are obtained from the time-diagonal elements of the cluster Green's function (opposed to the off-diagonal elements shown in figure \ref{gsiam}).
As expected physically, the strong impurity polarization dissipates into the rest of the system and decreases with time. 
However, the system is finite and small which causes a strong revival of the moment at a time $\approx 10$. 
For $t \approx 5$ one can see the polarization to be at a maximum at $i=8$, i.e.\ at the opposite edge of the chain.

A maximum time of the order of $t_{\rm max}=10$ is dictated by the CPT due to the necessity to solve the CPT equation by time discretization and inversion of matrices in $t,t'$ (see also below).
This limitation of the CPT to the short-time physics is actually characteristic for any nonequilibrium cluster-embedding method although a somewhat larger $t_{\rm max}$ is possible using advanced quadrature formulas.
For $t_{\rm max}=10$ the calculations can be done by constructing the Krylov space only once without any restart.
Namely, as can be seen by comparing the results for different Krylov-space dimensions $n$ in figure \ref{mag}, convergence is obtained for $n \approx 50$ in this example, i.e.\ for rather moderate values.

%*******************************************************************************
\begin{figure}[t]
\centerline{\includegraphics[width=0.8\columnwidth]{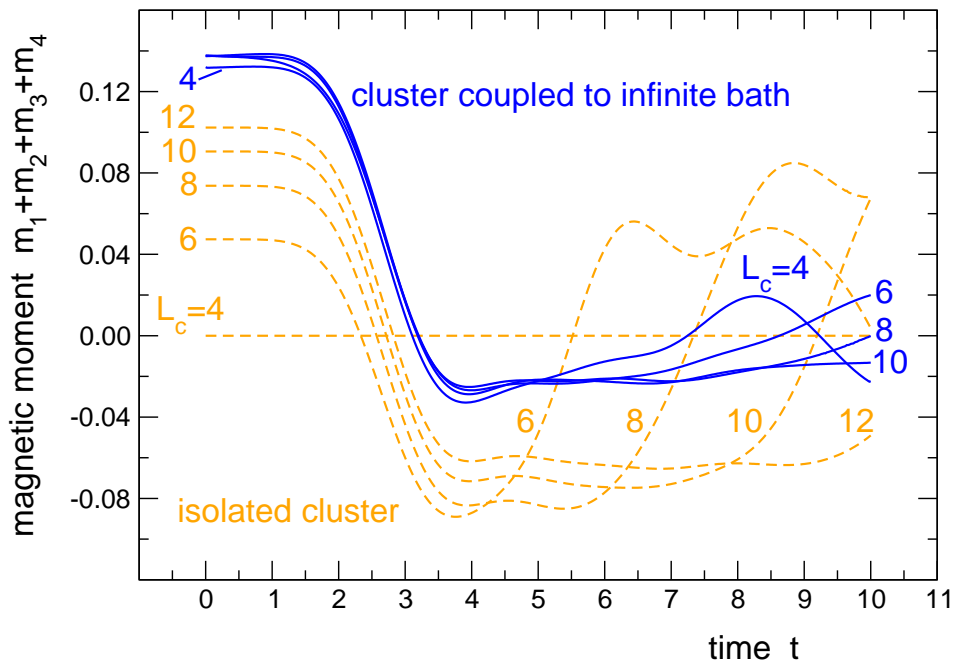}}
\caption{ 
Time dependence of the polarization of the first four sites for the model displayed in figure \ref{siam}.
Results for isolated ($V=0$) clusters of different size $L_c$ (dashed lines) and for the full ($V=T$) system (solid lines) as obtained by NE-CPT for $L_c=4 - 10$ at $U=T=1$. 
}
\label{fss}
\end{figure}
%*******************************************************************************

With the contour-ordered Green's function of the isolated cluster and the bath Green's function at hand, i.e.\ with $\ff G'$, the Green's function of the full model $\ff G$ can be obtained from (\ref{cpteq}). 
According results are shown in Figure \ref{fss}.
For the calculation, we have considered the SIAM with a cluster of $L_c=4-12$ sites and a bath consisting of $L_b=1000$ sites simulating a semi-infinite system (see figure \ref{siam}). 
The full contour-ordered Green's function $G_{ij\sigma}(t,t')$ for sites $i,j$ in the cluster is obtained by solving the time-discretized CPT equation (\ref{cpteq}). 
For the Keldysh branch along the real time axis from $t=0$ up to $t=t_{\rm max}=10$, a time integration step of $\delta t = 0.02$ has turned out to be sufficient for convergence.
For the Matsubara branch along the imaginary axis we find converged results for $\beta = 15$ and $\delta \tau = 0.02$. 
Note that not only the time evolution of the final state but also the initial equilibrium state is treated by means of the CPT.
Therefore, to take potential-scattering vertices into account for the initial state, the Matsubara branch must be included in the calculation. 
For further technical details on the NE-CPT and the solution of the CPT equation exploiting symmetries, we refer to \cite{BP11}.

For the discussion of the results, consider the initial state at $t=0$ first.
The magnetic field $h=0.2$ applied at the impurity site causes a sizable impurity polarization $m_1=n_{1\uparrow} - n_{1\downarrow} \approx 0.24$ (not shown). 
For the half-filled system, strong antiferromagnetic spin correlations then lead to a polarization cloud close to the impurity with alternating local magnetic moments $m_i$ which decrease in size with increasing $i$.
The figure shows the net polarization at the first four sites of the semi-infinite system which amounts to $\sum_{i=1}^4 m_i \approx 0.137$ in the initial state. 
Note that the convergence with increasing cluster size $L_c$ is extremely fast for the CPT results (full blue lines at $t=0$) as compared to the results for an isolated cluster (dashed orange lines at $t=0$). 
For the isolated cluster with $L_c=4$ the total polarization even vanishes as the ground state is a singlet for the considered small $h$. 

The finite net polarization close to the impurity is expected to dissipate into the infinite, initially unpolarized bath.
Indeed, the CPT shows that as a function of time the polarization relaxes quickly, overshoots a bit and then slowly approaches the equilibrium value, i.e.\ vanishes.
Comparing the different cluster sizes, we can say that the almost exact final-state dynamics is obtained with $L_c=10$ for times up to $t_{\rm max}=10$.
This is traced back to the fact that due to the spatial separation between $U$ and $V$ vertices, vertex corrections decrease with increasing distance $L_c$ and are sufficiently small for $L_c=10$. 
Opposed to the CPT, the results for the isolated cluster exhibit a strongly oscillatory behavior as a function of time, and finite-size scaling is obviously impossible for larger times. 

\section{Summary}
\label{con}

For the calculation of dynamical correlation functions by means of Krylov-space techniques, it makes a big difference whether the frequency-dependent Green's or spectral function is addressed by approximating the resolvent of the Hamiltonian, or, on the other hand, the time-dependent correlation function by approximating the exponential of the Hamiltonian, followed by a Fourier transformation to frequency representation. 

We first summarize the results for the equilibrium (zero-temperature) single-particle Green's function. 
In the first case, the replacement $(\omega - H)^{-1} \ket{i_0} \mapsto P (\omega - H)^{-1} P \ket{i_0}$ represents an approximation that conserves the first $n$ moments of the spectral density if $P$ is the projection onto an $n$-dimensional Krylov space:
Namely, the moments are related to the coefficients in the high-frequency expansion of the Green's function, and this is obtained via the expansion of the resolvent, $1/(\omega - H) = \sum_{r=0}^\infty H^r/\omega^{r+1}$. 
Then, conservation of the moments results from the fact that, by construction, $H^r \ket{i_0}$ can be exactly represented in the Krylov space $\ca K_n(\ket{i_0})$ if $r < n$.

In the second case, the replacement $\exp(-i H t) \ket{i_0} \mapsto P \exp(-iHt) P \ket{i_0}$ is numerically exact for a sufficiently short propagation time $t$. 
Using several restarts of the Krylov construction, this can be exploited to get the time-dependent Green's function and finally, by mean of fast Fourier transformation, the frequency-dependent Green's function and spectral density. 
This time-dependent Lanczos algorithm reduces to the conventional one if no restart at all is considered. 
Like the correction-vector method, it provides the numerically exact result at the cost of a largely increased numerical effort that is roughly proportional to the number of restarts $f$.
A too small $f$, however, leads to spectral densities violating causality. 
Comparing the conventional with the time-dependent Lanczos technique, the most significant deviations are found at the highest frequencies in the spectrum since the convergence of ground state and the low-lying excited states with increasing $n$ is very fast in the conventional method.
Generally, the approximation of the resolvent is a severe one if compared with the approximate Lanczos determination of the ground state.
On the other hand, due to the substantially larger CPU times necessary, one might tolerate this approximation, in particular in the context of the DMFT or dynamical cluster-embedding schemes where the error due to the finite small number of sites in the effective impurity or cluster model can be more severe.

For the nonequilibrium case, however, an approach based on the approximation of resolvents appears as rather ineffective:
Since single-particle correlation functions are no longer homogeneous in time, Fourier transformation to the frequency representation does not help in solving Dyson's equation which is central to any nonequilibrium dynamical embedding method.
Nevertheless, the contour-ordered Green's function can be represented in terms of resolvents, see (\ref{glesseromega}).
Treating these by projection onto appropriate Krylov spaces which necessarily must be constructed from the {\em excited} states as initial states, represents a much less controlled approximation compared with the equilibrium case.
On the other hand, a correction-vector method would provide the numerically exact result, but at the cost of the necessity to solve large sparse linear systems of equations for each {\em pair} of frequencies on a sufficiently dense frequency mesh. 
Apart from that, a three-dimensional Fourier back transformation is required in the resolvent-based approach. 

We have therefore suggested a four-step procedure to compute the Green's function as a function of $t$ and $t'$ on the Keldysh-Matsubara contour which is numerically exact and much more efficient than an approach to resolvents based on correction vectors:
After (i) finding the initial state $\ket{0}$ of the system as the ground state of an initial Hamiltonian by means of a standard Lanczos procedure, (ii) $\ket{\Phi(t)} = \exp(-iHt) \ket{0}$ is computed by means of the Krylov technique, followed by (iii) the back propagation $\ket{\Psi_\alpha(t)} = \exp(iHt) c_\alpha \ket{\Phi(t)}$ for every $t$ and (iv) the evaluation of a scalar product $\braket{\Psi_{\alpha'}(t')}{\Psi_\alpha(t)}$ from which the different components of the contour-ordered Green's function are obtained.
Step (iii) is most CPU time consuming.
Compared to the time-dependent Lanczos approach to the equilibrium Green's function, the main complication consists in the fact that the initial state $c_\alpha \ket{\Phi(t)}$ for the time propagation in (iii) is time-dependent itself. 
There are no problems, however, to get all components of $G_{\alpha\alpha'}(t,t')$ for a half-filled Hubbard or single-impurity Anderson model with $L_c=12$ sites and for $t,t' \lesssim 100$ on the real branches, for example, using a standard PC using the total particle number and the $z$-component of the total spin as good quantum numbers.

As a simple application of the nonequilibrium exact-diagonalization solver, we have considered the nonequilibrium cluster-perturbation theory \cite{BP11}.
For the single-impurity Anderson model in a semi-infinite chain geometry, a magnetic excitation that is localized in the vicinity of the correlated impurity is expected to dissipate into the uncorrelated and unpolarized bath in the process of time.
This is nicely seen within the NE-CPT if the hopping $V$, linking the linear cluster of the first $L_c$ sites with the infinite uncorrelated bath, is treated as the inter-cluster hopping by means of all-order perturbation theory in $V$ and $U$.
Within the NE-CPT the neglected vertex corrections are controlled by the spatial distance between the $U$ and the $V$ vertex and thus by the cluster size. 
Our calculations for $L_c = 4-10$ sites show a systematic improvement and give the essentially exact result on a time scale of $t \lesssim 10$ (in units of the inverse hopping).
This is just the scale which is typically accessible by means of dynamical impurity or cluster embedding approaches and which is relevant, for example, to estimate the speed of information processing in atomic-scale all-spin–based devices \cite{KWCW11}.

There are several points that may be addressed in future studies:
In case of driven systems or within the context of nonequilibrium DMFT, where the time dependence of the Hamiltonian $H(t)$ is more complicated than a simple sudden quench of a parameter, the Krylov time evolution must be carried out using a time discretization step $\Delta t$ that is considerably shorter than any characteristic time scale of $H(t)$.
More efficiently, higher-order commutator-free exponential time-propagation algorithms \cite{AF11} can be applied.
Another interesting line is the computation of {\em higher-order} Green's functions using the Krylov approach. 
Correlation functions depending on four independent time variables are required, for example, in the context of the nonequilibrium dual-fermion approach \cite{JLB+11}.
Finally, for the application of the nonequilibrium exact-diagonalization solver within the NE-CPT the use of advanced quadrature formulas is promising to extend the Keldysh branch, i.e.\ the limit $t_{\rm max}$ up to which observables can be traced.
Applications to two-dimensional lattice models are particularly interesting as there is hardly an alternative to an exact-diagonalization solver.

\ack
We would like to thank Ph.\ Jurgenowski for critically reading the manuscript.
The work is supported by the Deutsche Forschungsgemeinschaft 
within the Sonderforschungsbereich 925 (project B5)
and 
within the Sonderforschungsbereich 668 (project B3).

\end{document}